\documentclass{pasj00}
\begin{document}
\SetRunningHead{K. Watarai}{Light Curves from Eclipsing Black Hole Binaries}
\Received{2009/11/2}%{yyyy/mm/dd}
\Accepted{2010/02/16}%{yyyy/mm/dd}

\title{Optical Light Curves of Luminous Eclipsing Black Hole X-ray Binaries}
%\title{Optical Light Curves from Eclipsing Black Hole X-ray Binaries}

%%% begin:list of authors
%\author{Ken-ya \textsc{Watarai} %\altaffilmark{1,2}%
%  \thanks{Research Fellow of the Japan Society for the Promotion of Science}}
%\affil{Astronomical Institute, Osaka Kyoiku University,
%Asahigaoka, Kashiwara, Osaka, Japan 582-8582}
%\affil{Research Fellow of the Japan Society for the Promotion of Science}
%\email{watarai@cc.osaka-kyoiku.ac.jp}

%%% begin:list of authors
% Do NOT capitalize all letters in "textsc".
\author{Ken-ya \textsc{Watarai} %
%  \thanks{Present Address is xxxxxxxxxx}
}
\affil{Kanazawa University Senior High School,
 Heiwa-machi, Kanazawa, Ishikawa, Japan 921-8105}
\email{watarai@kfshs.kanazawa-u.ac.jp}
\and
\author{Jun {\sc Fukue}}
\affil{Astronomical Institute, Osaka Kyoiku University,
Asahigaoka, Kashiwara, Osaka, Japan 582-8582}
%\email{fukue@cc.osaka-kyoiku.ac.jp}
%%% end:list of authors

%% `\KeyWords{}' always has to be placed before `\maketitle'.
 %Do NOT move this preamble from here!

\maketitle

\begin{abstract}
We examine optical $V$-band light curves in luminous eclipsing black hole X-ray binaries, using a supercritical accretion/outflow model that is more realistic than the formerly used ones.
In order to compute the theoretical light curve in the binary system, we did not only apply the global analytic solution of the disk, 
 but also include the effect of the optically thick outflow. 
We found that the depth of eclipse of the companion star by the disk changes dramatically when including the effect of the outflow. Due to the effect of outflow, we could reproduce the optical light curve for typical binary parameters in SS433. 
Our model with an outflow velocity $v \sim 3000 {\rm km/s}$ could fit whole shape of the averaged $V$-band light curve in SS433, but we found a possible parameter range consistent with observations, such as $\dot{M} \sim$ 5000--10000 $L_{\rm E}/c^2$ (with $L_{\rm E}$ being the Eddington luminosity and $c$ being the speed of light) and $T_{\rm C}$=10000K--14000 K for the accretion rate and donor star temperature, respectively. 
Furthermore, we briefly discuss observational implications for ultraluminous X-ray sources. 
\end{abstract}

\section{Introduction}
The profile of a light curve in an astronomical object includes a wide range of information, i.e., the spatial distribution and dynamical motion of gas, or its radiative processes, etc. 
The analysis of the light curve has been therefore known as an important tool from the old days. 
Especifically, in the case of the eclipsing binary system, which has a compact star, it is useful to suppose the brightness distribution of the accretion disk around the compact star. 
This method can also be applied to black hole candidates.
Generally, it is difficult to discern light from the companion star and light from the accretion disk, but if the object has an eclipse,
 it may be possible for two ingredients to separate. 
Moreover, black hole candidates show the (partially) eclipsing property, 
and we could close in on the feature of the accreting gas,
 or the black hole mass, or spin, etc (Fukue 1987; Watarai et al. 2005; Takahashi \& Watarai 2007). 

For the last decade,
 a large number of luminous black hole candidates (BHCs)
 have been found in nearby galaxies. 
They are called ``ultraluminous X-ray sources (ULXs)'', but their origin is still unknown (Makishima et al. 2000; Roberts et al. 2005). 
X-ray data analysis is the mainstream study of the ULXs,
 but other band data will be useful for evaluating other binary parameters. 
It is no wonder that eclipsing binaries are detected among the BHCs. 
In fact, several eclipsing black hole binaries have been discovered (Orosz et al. 2007 in M33; Ghosh et al. 2006 in NGC4214). 

To examine the observational properties of such luminous eclipsing binaries,
 light-curve fitting will be a powerful tool. 
The conventional study of a light curve in a binary system
 applies a geometrically thin disk, a so-called standard accretion disk
 (Shakura \& Sunyaev 1973). 
This model may have applicability to the high/soft state in black hole candidates, but it may not extend to the low/hard state, very high state, or super-critical state, which is a more luminous state than the high soft state. 
Luminous black hole candidates that exceed to the Eddington luminosity seem to accrete a large amount of gas via a companion star, and the disk becomes geometrically thick. This type of accretion flow is called a ``slim disk'' (Abramowicz et al. 1988; Watarai et al. 2000). 
The geometrical thickness of this disk can play an important role in covering a companion star, depending on an angle, and contribute to changing the shape of the light curve. 
It is necessary to include it in a calculation of bright binary system properly. Fukue et al. (1997, 1998) performed light curve analysis in SS433
 using geometrically thick torus model by Madau (1988),
 but it is known that the disk model is thermally unstable. 
Hirai \& Fukue (2001) compared the optical light curve in SS433 
 with (thermally stable) supercritical accretion disks,
 but they adopted the self-similar solutions even at the disk outer region. 
%As a result, their calculation overestimates the scale height and effective temperature of the disk. 
We thus adopted a thermally stable, more realistic (appropriate) treatment of the disk model, and compared the model with observations of super-critical accreting objects such as SS433 and ultraluminous X-ray sources. 

In addition,
in the previous studies
the mass outflow from a supercritical disk was not included,
but a naked supercritical disk was considered.
Hence, in the present study
we consider the supercritical accretion disk
with an optically thick outflow from the center.

In the next section, we introduce the assumptions of our model. 
In section 3, we briefly show the light curve calculation method. 
Calculation results are presented in section 4. 
The final section is devoted to concluding remarks. 

\section{Model for Accretion and Outflow under Supercritical Accretion Flows}

In this section, we calculate the light curve when the binary star system that fills Roche-lobe is assumed, and a supercritical accretion has occured in the compact star. 

Ideas of supercritical accretion have been proposed by many authors in early 1970's (Shakura \& Sunyaev 1973; Abramowicz et al. 1978;
 Jaroczynski et al. 1980; Paczy\'nski \& Wiita 1980). 
Whether accretion that exceeds the Eddington luminosity is possible or not has been doubted for many years. 
The solution for supercritical accretion had already been obtained in one dimension (Abramowicz et al. 1988). 
Due to the development of the computer, supercritical accretion is actually known to be reproduced by numerical simulations (Okuda 2002; Ohsuga et al. 2005; Ohsuga \& Mineshige 2008). 
It thus becomes impossible for us to deny supercritical accretion any longer. 
In the next subsections, we briefly introduce our model and assumptions. 

\subsection{Model for Supercritical Accretion Flows}

The photon trapping radius characterizes supercritical accretion
 at the radius where advective energy transport becomes important (Begelman \& Meier 1982; Ohsuga et al. 2002). 
This radius accords with the radius that the gravity of the disk balances with the radiation pressure, and it has been suggested that an outflow may originate inside it (Shakura \& Sunyaev 1973; Lipunova 1999; Fukue 2004; Heinzeller \& Duschl 2007). 
%Mass loss effect will be important inside the photon trapping radius. 

However, the outflow blows a very tiny area compared with the size of the whole disk. It actually will not have an influence on computing optical flux. 
Recently Takeuchi et al. (2009) analyzed 2D RHD simulation results by Ohsuga et al. (2005), and rebuilt a one dimension model, but the effective temperature distribution of the one dimension model hardly changes. 
Therefore, we can use an one-dimension model safely without any outflow effect.

\vspace{0.5cm}
\subsubsection{Radiation-pressure dominated regime: $\kappa_{\rm es} \gg \kappa_{\rm ff}$}

For radiation-pressure dominated regime, Watarai (2006) constructed analytical formulae that can be applied for a wide range of accretion rates, and it has been shown that these solutions are a good approximation of the numerical solutions. 
According to Watarai (2006), the scale height of the disk is given by 
\begin{equation}
H_{\rm a} =  3.0 f(\hat{r},\dot{m})^{1/2} \hat{r}.
%H =  \sqrt{(2N+3) \frac{ B \Gamma_{\Omega} \Omega_0^2}{\xi}} f(\hat{r},\dot{m})^{0.5} r.
\label{eq:ha}
\end{equation}
This solution is characterized by the ratio of the advective cooling rate to the viscous heating rate, i.e., $f=Q_{\rm adv}^-/Q_{\rm vis}^+$, which can be represented by an analytical form dependent on the radius and the mass accretion rate, $f(\hat{r},\dot{m})$. 
The radius $\hat{r}$ is normalized by the Schwarzschild radius, 
 and the $\dot{m}$ represents the mass accretion rate in Eddington units
 ($\dot{M}_{\rm Edd}=L_{\rm E}/c^2$). The explicit form of $f(\hat{r},\dot{m})$ is given by 
\begin{equation}
f(\hat{r},\dot{m}) = \frac{1}{2} \left[ D-2 (\hat{r}/\dot{m})^2 +2 -D (\hat{r}/\dot{m}) \sqrt{D^2 (\hat{r}/\dot{m})^2+4} \right] 
\end{equation}
where $D$ is a constant of order unity
 (e.g., $D \approx 2.18$ for a polytropic index $N=3$). 
The function $f(\hat{r},\dot{m})$ is close to unity for an advection dominated regime, and it is close to zero for a radiative cooling dominated regime (see Watarai 2006 for more details). 

The effective temperature distribution is given by 
\begin{equation}
T_{\rm eff} \approx 4.48 \times 10^7 f(\hat{r},\dot{m})^{1/8} m^{-1/4} \hat{r}^{-1/2} {\rm K}.
%T_{\rm eff} \approx 2.52 \times 10^7 f(\hat{r},\dot{m})^{1/8} \left(\frac{m}{10}\right)^{-1/4} \hat{r}^{-1/2} {\rm K}.
\end{equation}

The boundary radius between the radiation-pressure dominated regime and the gas-pressure dominated regime is located at 
\begin{eqnarray}
\hat{r}_{\rm rad-gas} &=& 18 (\alpha m)^{2/3} \dot{m}^{16/21} \\
 &\approx& 601 (\alpha/0.1)^{2/3} (m/10)^{2/3} (\dot{m}/100)^{16/21}. 
\end{eqnarray}
The same equations are posed by Shakura \& Sunyaev (1973). 
We note that analytic solutions shown here are useful for radiation-pressure dominated regime. 

\vspace{0.5cm}
%\subsubsection{Gas-pressure dominated regime: $\kappa_{\rm es} \gg \kappa_{\rm ff}$}
\subsubsection{Gas-pressure dominated regime}

In the gas-pressure dominated regime, if electron scattering dominates the
 opacity, the scale height of the disk and the effective temperature distribution are 
%For gas-pressure dominated regime, electron scattering dominated for its opacity, the scale height of the disk is given by 
\begin{equation}
H_{\rm b} = 2.7 \times 10^3 \alpha^{-1/10} m^{9/10} \dot{m}^{1/5}
 \hat{r}^{21/20},
\label{eq:hb}
\end{equation}
\begin{equation}
T_{\rm eff} \approx 3.50 \times 10^7 m^{-1/4} \dot{m}^{1/4} \hat{r}^{-3/4} {\rm K},
\label{eq:teff}
\end{equation}
with the same formula by Shakura \& Sunyaev (1973). 

The transition radius where $\kappa_{\rm es} \sim \kappa_{\rm ff}$ is 
%The boundary radius between the $\kappa_{\rm es}$ dominated regime and the $\kappa_{\rm ff}$ dominated regime is 
\begin{equation}
\hat{r}_{\rm gas, out} = 2.5 \times 10^3 \dot{m}^{2/3}, 
\end{equation}
 and thus it is a simple function of the mass accretion rate. 

%\vspace{0.5cm}
%\subsubsection{Gas-pressure dominated regime: $\kappa_{\rm ff} \gg \kappa_{\rm es}$}
The outer region of the disk is dominated by free-free opacity, and the scale height is given by 
%The outer region of the disk is a free-free opacity dominated regime. 
%The scale height of the disk is given by 
\begin{equation}
H_{\rm c} = 1.5 \times 10^3 \alpha^{-1/10} m^{9/10} \dot{m}^{3/20}
 \hat{r}^{9/8},  
\label{eq:hc}
\end{equation}
 with the effective radial dependence of the temperature as in equation (\ref{eq:teff}).
%The effective temperature distribution is the same as equation (6), 
%which is proportional to $\hat{r}^{-3/4}$. 

We ignore the irradiation by the disk itself or the photosphere of the outflow,  because the irradiation dominated regime appears at the outer region of the disk, and the temperature and geometrical effects do not contribute to the optical light curves (less than 10 \%). 
To avoid confusion of the model, we decided to handle an easier model.

\subsection{Model for Massive Wind}

Shakura and Sunyaev (1973) proposed the massive (supercritical) outflow, which is formed by the strong radiation from the disk. 
The size of the photosphere surface made by the outflow becomes large and has an influence on the form of the light curve.
Thus, we should include the effect of the outflow in our model. 
We do not include the collimated, relativistic jet component in this paper. 

Here we introduce a simple wind model by Abramowicz et al. (1991), which assumes a spherical symmetry and uniform outflow velocity. 
The density of the wind, $\rho_{\rm w}$, is 
\begin{equation}
\rho_{\rm w} = \left(\frac{\dot{M}_{\rm out}}{4 \pi v \gamma} \right) R^{-2}
\end{equation}
where $\dot{M}_{\rm out}$ is the mass outflow rate, $v$ is the velocity of the gas, $\gamma$ is the Lorentz factor: $(1-\beta^2)^{1/2}$, where $\beta=v/c$,
 and $R$ is the distance from the black hole.  

The mass outflow rate should be determined by the physics of the interaction between the disk and outflow, i.e., $\dot{M}_{\rm out} = \eta \dot{M}_{\rm acc}$, where $\eta$ is the efficiency of the outflow gas from accreting gas. 
In our present study, we assume $\eta=1.0$ for simplicity. 
That is, our model simply assumes that all accreting matter changes to the outflow at the disk inner edge ($3r_{\rm g}$). 
The location of the boundary layer between the outflow and the disk may have a strong impact on the emerging X-ray spectrum, but it is not expected to have an enormous  influence on the optical band, since the optical emitting region of the photosphere is far away from the disk inner edge. 

\section{Binary Light Curve Calculation}

To calculate the $V$-band flux in a binary system,
 we adopt the ``Ray-Tracing Method''. 
We suppose that the binary star fills the Roche lobe and transports its mass to the compact star. 
The shape of the companion star reflects the shape of the potential. 
%The surface of the disk and the secondary star of each binary phase
% is calculated by numerical integrations. 
The photon propagates from an emitting point on the surface of the disk or that of the photosphere of the wind to the distant observer. 
According to Fermat's principle, however, the light rays are traced
 from the observer's display coordinate $(x_{\rm s},y_{\rm s},z_{\rm s})$. 
After the ray arrives at the surface of the disk/wind, 
 we evaluate the geometrical thickness and effective temperature 
 of the disk/wind model presented in the previous section. 
The observed flux is integrated in each optical bands ($V$, $R$, $I$),
 assuming the blackbody radiation at the surface of the disk, star, and photosphere of the outflow. 

The spatial resolution of the calculation is about %$0.01 a - 0.001 a$, 
$1 \% - 0.1 \%$ of the binary distance, 
 which equals to $10^4 r_{\rm g}$ - $10^3 r_{\rm g}$. 
This resolution is sufficient to resolve optical flux from the binary system. 
%We can roughly estimate the enough mesh size to compute the optical flux in following way. 
%Using Wien's displacement law, we can calculate the maximum temperature $T_{\rm opt}$ for optical frequency ($T_{\rm opt} = \nu_{\rm opt}/ 5.88 \times 10^{10} = 1.6 \times 10^6$ [K]). 
%If we regard the $T_{\rm opt}$ as the disk temperature, the emission seems to come from the radius  $\sim 1.2 \times 10^4 r_{\rm g}$ by the standard accretion disk model. 
Thus, mesh size of this calculation is at least much smaller than 10 \% of this radius. 

\subsection{Location of the Photosphere}
Location of the wind photosphere is estimated from the point where the optical depth measured from a distant observer equals to unity, that is, 
\begin{equation}
\tau_{\rm ph} = 1 = \int_{R_{\rm ph}}^{\infty} \gamma (1-\beta \cos{\theta}) \kappa \rho_{\rm w} ds.
\label{eq:tauph1}
\end{equation}
where the $R_{\rm ph}$ is location of the last scattering surface
 of the photon in the disk coordinates, $\gamma$ is the Lorentz factor, $\kappa$ is the opacity for electron scattering,
 $\beta$ is the velocity of the wind normalized by the speed of light $c$,
 and the $\theta$ is the inclination angle.
Abramowicz et al. (1991) obtained an analytic solution of a moving plasma,
 and the shape of the photosphere in their model does not have spherical symmetry. We use their formulae in this paper. 

In some cases, analytic formulae derived by King (2003)
 are useful to estimate the size of the photosphere. It is given by 
\begin{equation}
% R_{\rm ph} \sim \frac{\kappa c \dot{M}_{\rm out}}{v \dot{M}_{\rm crit}} 
% = \frac{\kappa  \dot{m}_{\rm out}}{\beta}
 R_{\rm ph} \sim \frac{\kappa \dot{M}_{\rm out}}{4 \pi b v}, 
\end{equation}
 where $R_{\rm ph}$ is the radius of the photosphere, $\kappa$ is the opacity for electron scattering, and $b$ is the eccentricity of the photosphere. The value of $b$ is $0.5 \to 1$, which is fixed at unity throughout this paper. 

In this paper, we assume that the mass outflow rate, $\dot{M}_{\rm out}$, is equal to the mass accretion rate at the central region. 
The mass outflow rate in our model gives the maximum rate. 
Typical size of the photosphere is given by 
\begin{equation}
 \frac{R_{\rm ph}}{r_{\rm g}} = \frac{\dot{m}_{\rm out}}{2 \beta}
 = 10^5 \left(\frac{\dot{m}_{\rm out}}{2000}\right) \left(\frac{\beta}{0.01}\right)^{-1}. 
\label{eq:rph2}
\end{equation}
As we show later, if the size of the photosphere is much smaller
 than the disk size, the photosphere does not influence the shape of the eclipsing light curves. 

\subsection{Temperature and Luminosity at the Photosphere}
%It is difficult to determine the temperature on the photosphere. 
We evaluate the maximum temperature of the photosphere by using the following procedure. 
First, we estimate the size of the photosphere of the outflow. 
Assuming that photons ejected from the disk surface inside the photosphere is conserved until escaping photons from the surface of the photosphere,
 i.e., all photons are generated inside the disk, not be generated in the wind. 
The temperature of the wind photosphere $T_{\rm w}$ then is given by 
\begin{equation}
\sigma T_{\rm w}^4 \approx \frac{L_{\rm disk}}{4 \pi (bR_{\rm ph})^2}. 
\label{eq:tw}
\end{equation}
Here the luminosity of the disk, $L_{\rm disk}$, is given by
\begin{equation}
L_{\rm disk} = \int_{R_{\rm in}}^{R_{\rm ph}}  2 \sigma T_{\rm eff}^4 \cdot 2 \pi r dr. 
\end{equation}

To determine the location of the disk surface or the photosphere of the outflow, we integrated the optical depth from an observer at infinity to the surface of $\tau_{\rm ph}=1$ along the line of sight. 
After the light rays arrived at the surface, we measured the temperature using equation (\ref{eq:tw}) and the geometrical thickness of the disk from the underlying disk/wind models. 

\section{Optical Flux Images}

\subsection{Case without Outflow }
Let us see the results when only a naked supercritical disk is considered. 
In figure 1, we show the $V$-band ($5.11 \times 10^{14}$ Hz -- $6.61 \times 10^{14}$ Hz) flux images at different binary phases. 
The mass accretion rates is $\dot{m}=10^3$, the inclination angle is $i=70^\circ$,
 mass ratio is $q=M_{\rm X}/M_{\rm C}=1$, temperature of the companion star is $T_{\rm C}=15000$ K, and velocity of the wind is $\beta=0$, respectively. 
As can be seen in figure 1, the geometrical thickness of the disk is thin compared with the size of the disk. 
It is understood that the thickness of the disk does not affect the light curve for $\dot{m}=10^3$. 
The scale height of the disk weakly depends on the $\dot{m}$ as can be seen equations (\ref{eq:ha}), (\ref{eq:hb}), (\ref{eq:hc}). 

Figure \ref{fig:nowind-i-lc} is the time-variation of $V$-band flux for various inclination angles. The mass accretion rate is set to be $\dot{m}=10^3$. 
When the inclination angle increases,
 the flux drops at phase 0.5. 
This is because the disk hides emission from a secondary star. 
Optical emission from the secondary star is much larger than that from the disk for $\dot{m}=10^3$. 

As mass accretion rate increases,
 the disk becomes as luminous as the companion star (see figure \ref{fig:nowind-md-lc}). 
Some previous studies applied a thick disk model to the optical light
curve analysis in SS433 (Sanbuichi \& Fukue 1993; Fukue et al. 1997, 1998;
 Hirai \& Fukue 2001). 
However, their model is called ``thick torus model''
(Abramowicz et al. 1978; Madau et al. 1988), and it is thermally/secularly unstable (Kato et al. 1998). 
Apart from the structure of the inner most region of the disk, our analytic disk model (i.e., slim disk model) is applicable to the supercritically accreting regime (Watarai 2006). The model can produce more plausible temperature and scale height. 
Hence, our model for a bright object is more realistic than that of past researche. 

\begin{figure}
  \begin{center}
    \FigureFile(65mm,65mm){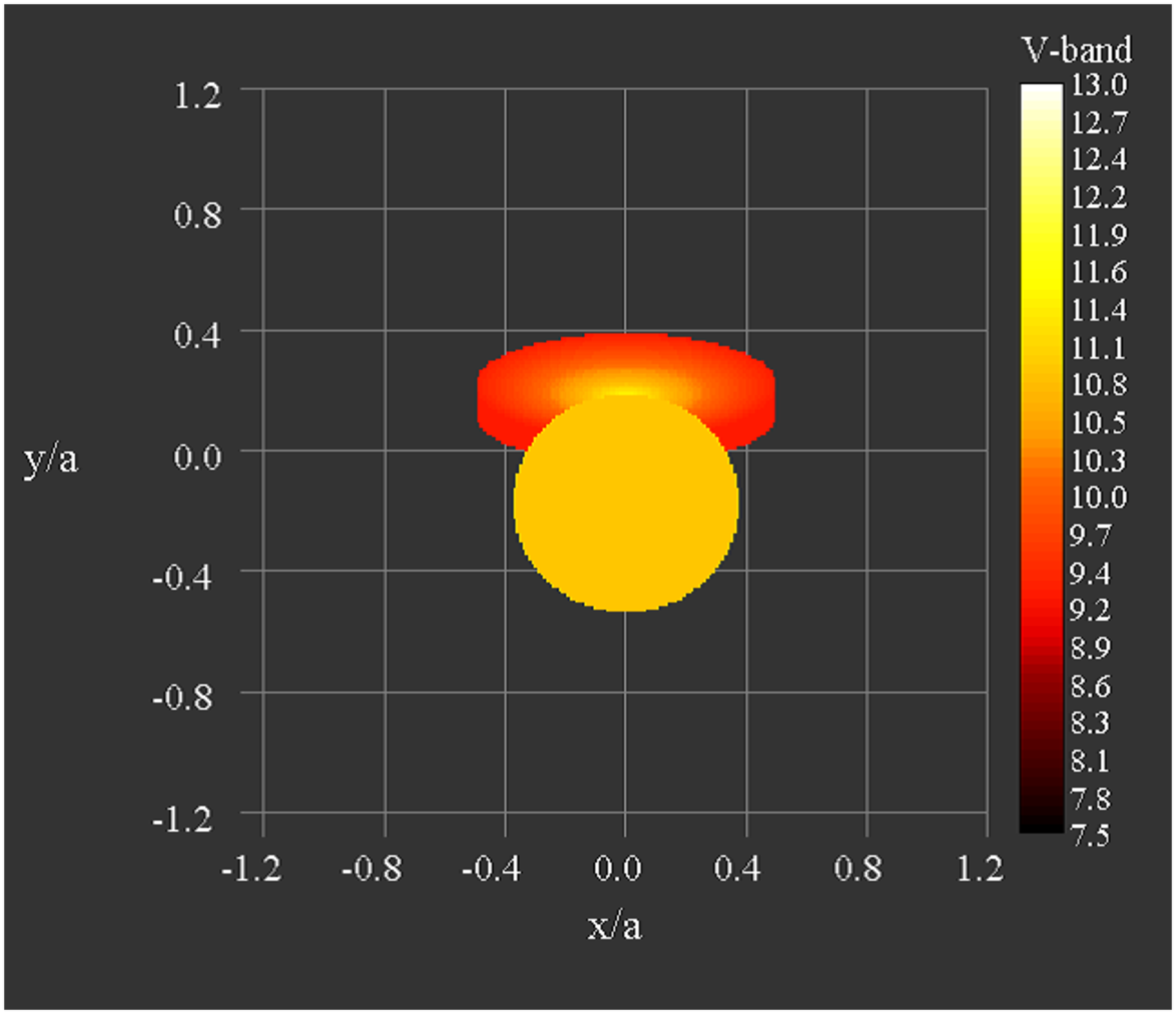}
    \FigureFile(65mm,65mm){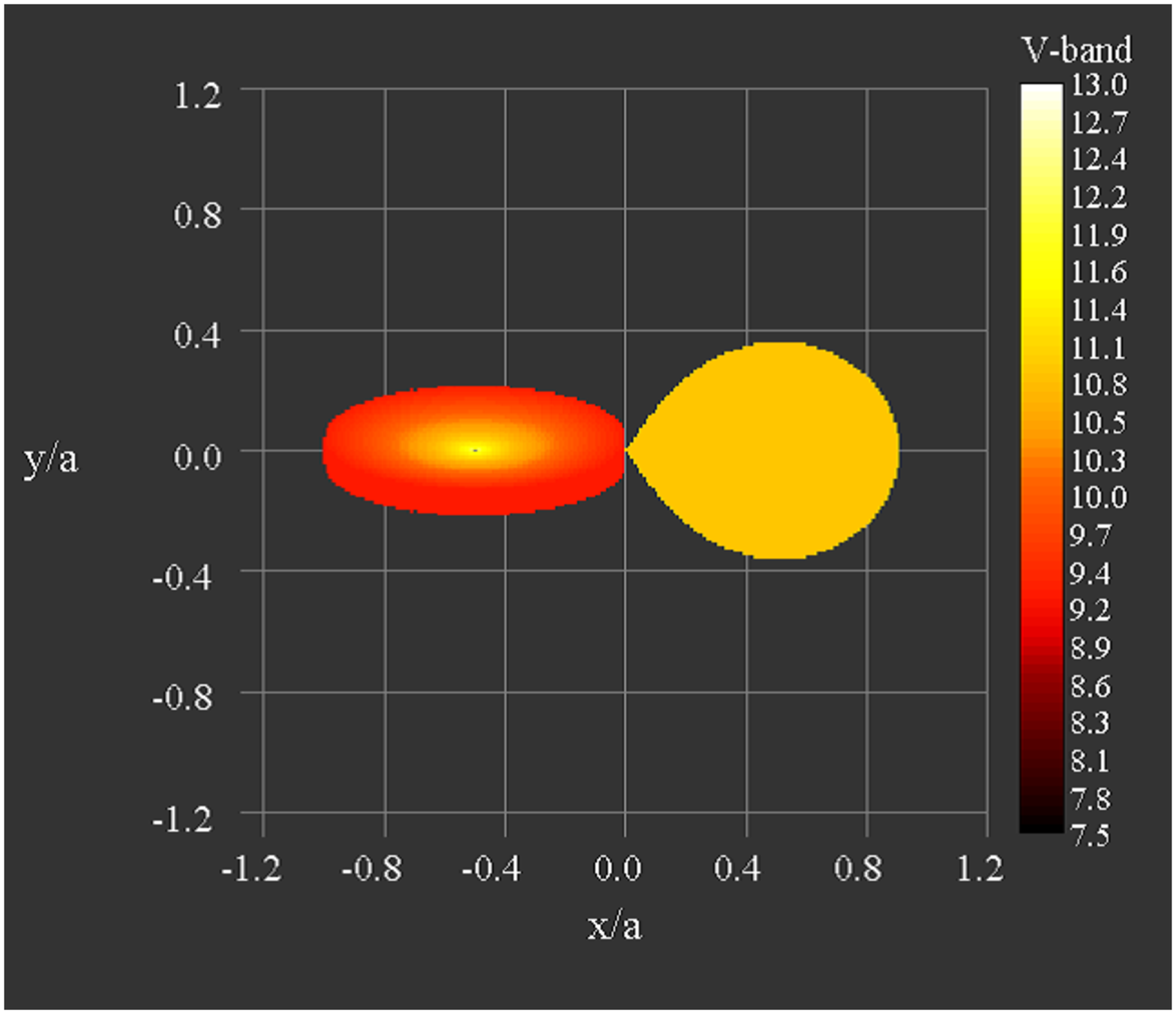}
    \FigureFile(65mm,65mm){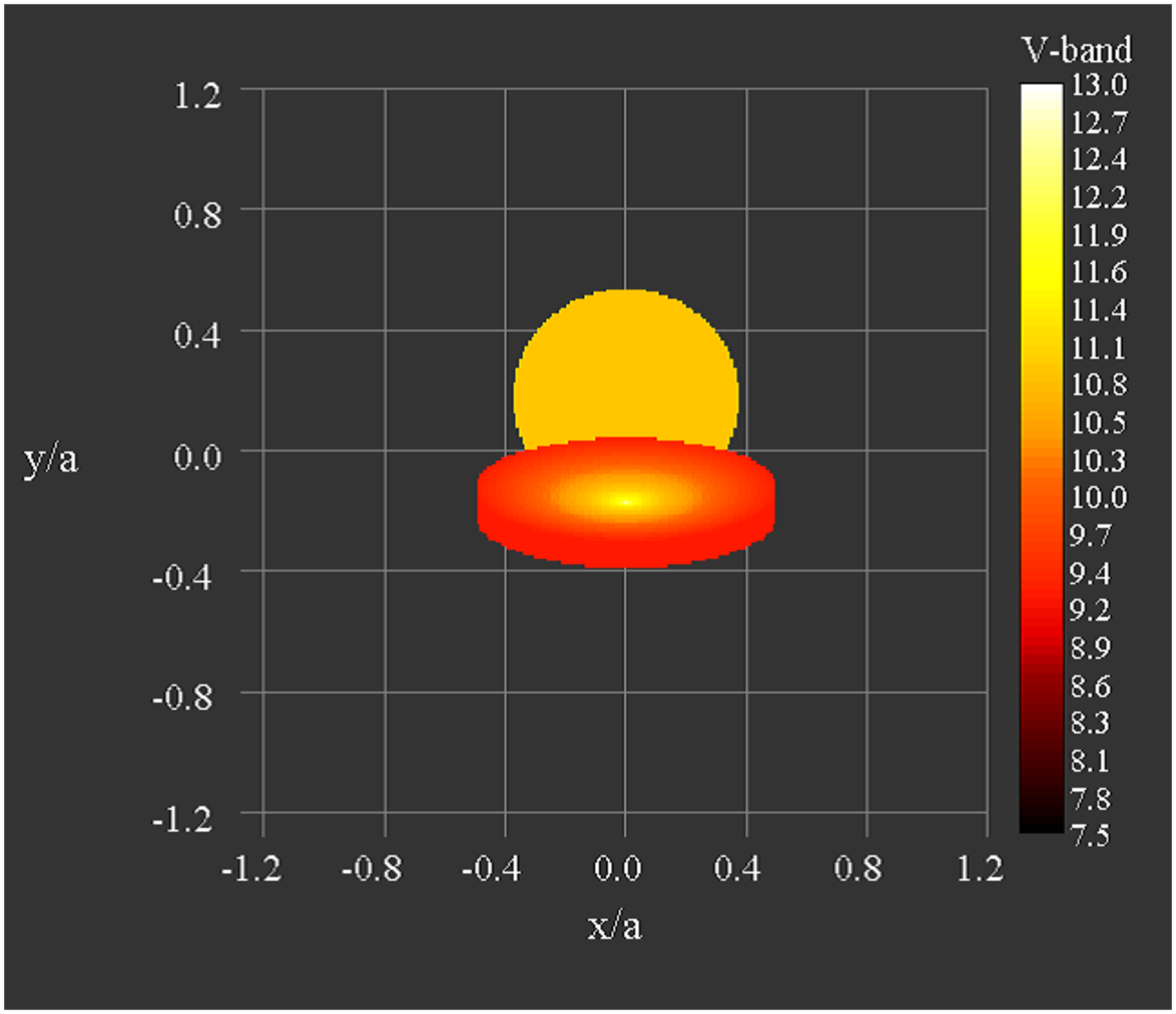}
    \FigureFile(65mm,65mm){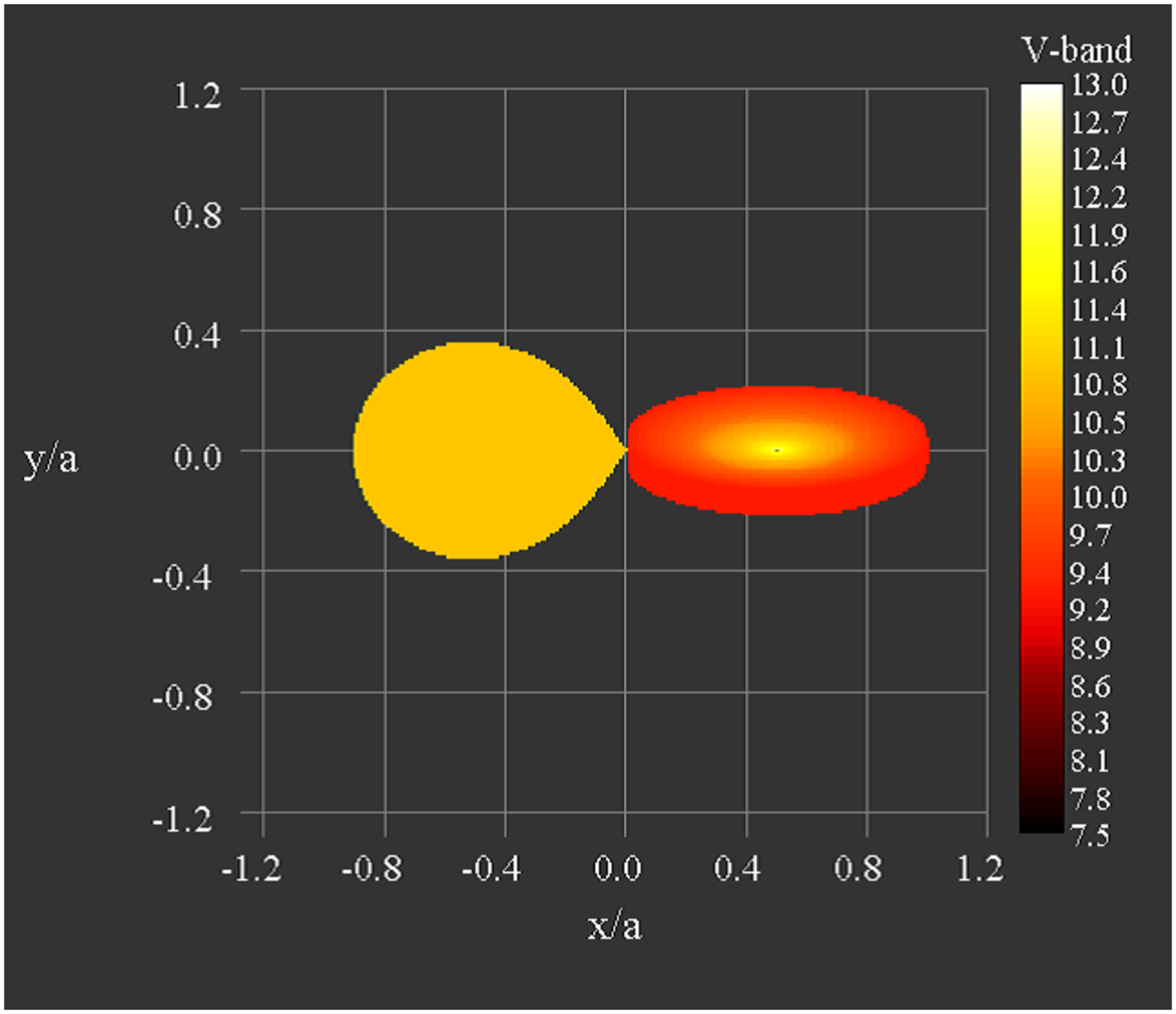} 
%    %%% \FigureFile(width,height){filename}
  \end{center}
  \caption{$V$-band flux images in no-outflow models with various phases (0, 0.25, 0.5, and 0.75). 
Horizontal and vertical axis are normalized by the binary distance $a$. 
The mass accretion rate is set to be $1000 \dot{M}_{\rm crit}$,
 and mass ratio $q=M_{\rm x}/M_{\rm c}$ is $q=1.0$. 
The inclination angle is $70^\circ$,
 and the temperature of the companion star $T_{\rm c}$ is 15000 K. }
\label{fig1}
\end{figure}

\begin{figure}
  \begin{center}
    \FigureFile(80mm,80mm){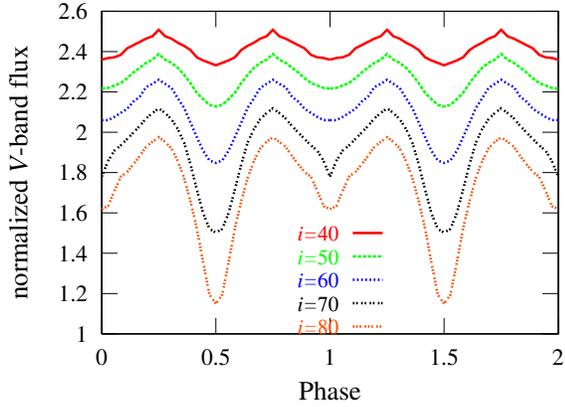}
  \end{center}
  \caption{Theoretical V-band light curves expected from our disk model (no wind) with different inclination angles. 
The inclination angles are $i=40^\circ,50^\circ,~60^\circ,~70^\circ$, and $80^\circ$ from bottom to top. 
Other parameters are $q=M_{\rm X}/M_{\rm C}=1.0$, $\dot{m}=10^3$, and $T_{\rm c}=15000 {\rm K}$, respectively. }
\label{fig:nowind-i-lc}
\end{figure}

\begin{figure}
  \begin{center}
    \FigureFile(80mm,80mm){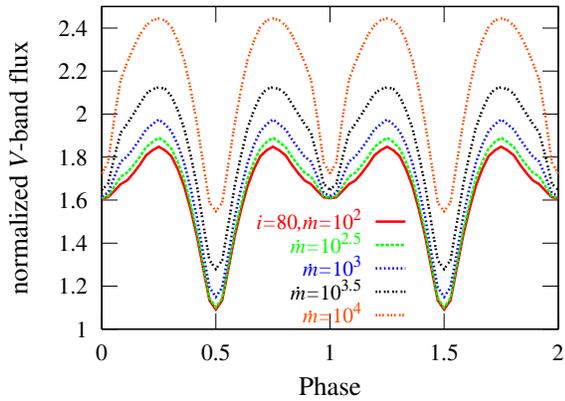}
  \end{center}
  \caption{Same as figure \ref{fig:nowind-i-lc}, but as a function of mass accretion rates (no wind). 
The mass accretion rates are $\dot{m}=10^2,10^{2.5},10^3,10^{3.5}$, and $10^4$ from bottom to top. 
Other parameters are $q=1.0$, $i=80^\circ$, and $T_{\rm c}=15000 {\rm K}$, respectively. }
\label{fig:nowind-md-lc}
\end{figure}

\subsection{Case with Massive Wind}

\begin{figure}
  \begin{center}
    \FigureFile(80mm,80mm){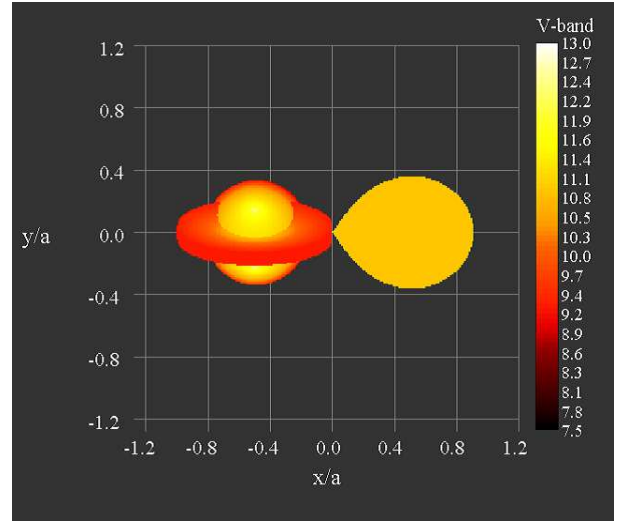}
    \FigureFile(80mm,80mm){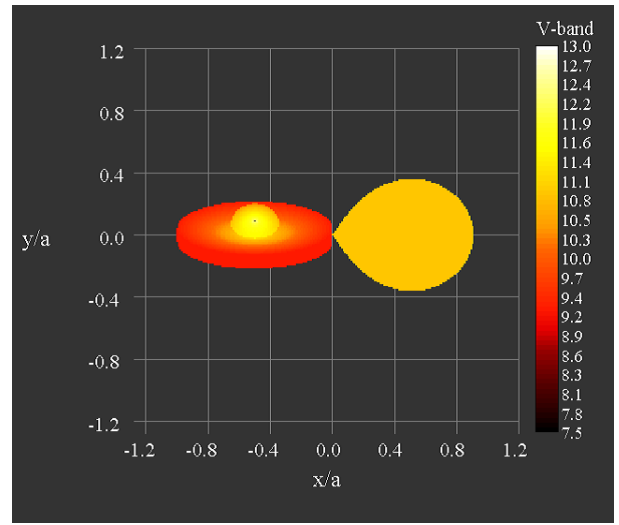}
  \end{center}
  \caption{ V-band images including the  effect of wind at phase 0.5 with various wind velocity $\beta$ = 0.003 (top) and 0.005 (bottom). 
Other parameters are $q=1.0$, $\dot{m}=10^3$, $i=80^\circ$, and $T_{\rm c}=15000 {\rm K}$, respectively. 
}
\label{fig:md1e3beta}
\end{figure}

 \begin{figure}
  \begin{center}
    \FigureFile(80mm,80mm){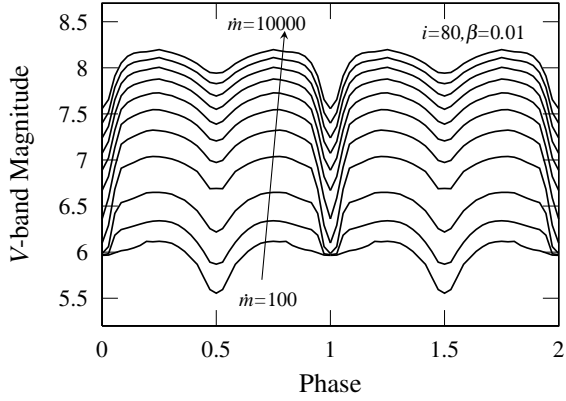}
    %%% \FigureFile(width,height){filename}
  \end{center}
  \caption{ V-band light curves with various mass accretion rates. The mass accretion rates are $\dot{m}$=100, 1000, 2000, 3000, 4000, 5000, 6000, 7000, 8000, 9000, and 10000 from bottom to top. 
Other parameters are $q=1.0$, $\beta$=0.01, and $i=80^\circ$. }
\label{fig:beta001}
\end{figure}

\begin{figure}
  \begin{center}
    \FigureFile(80mm,80mm){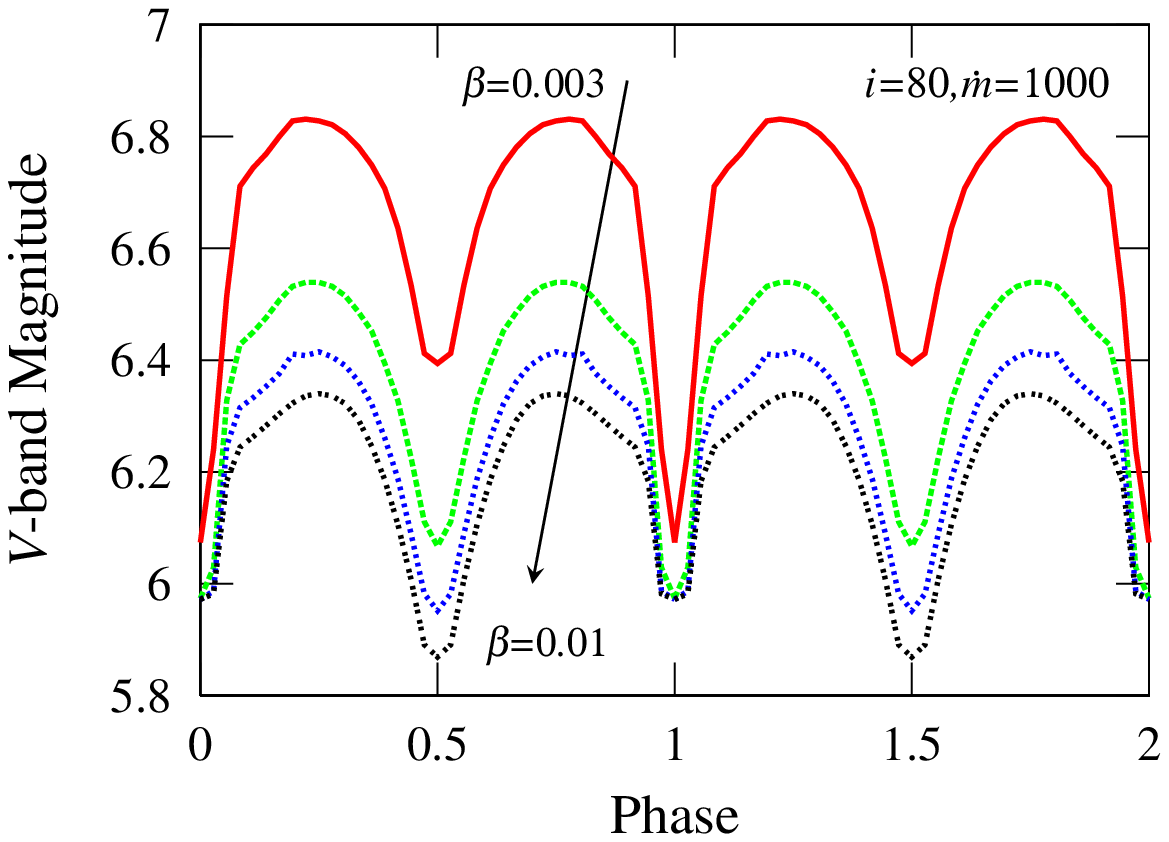}
    %%% \FigureFile(width,height){filename}
  \end{center}
  \caption{Same as figure \ref{fig:beta001}, but the dependence of the wind velocity $\beta$. 
The wind velocities are 0.003, 0.005, 0.007, and 0.01 from top to bottom.}
\label{fig:beta2}
\end{figure}

Figure \ref{fig:md1e3beta} represents flux images of a model with massive wind. 
The spherical structure at the central region of the disk is the photosphere by the massive wind. 
The brightness of the photosphere gradually changes from the central region to the outer region. This is called the limb-darkling effect. 

Figure \ref{fig:beta001} shows V-band light curves for high inclination angle $i=80^\circ$. 
The most remarkable feature is that a primary minimum and a secondary minimum inversion  happen as the mass accretion rate increases. 
The $V$-band magnitude at phase 0 or 1 is larger than the magnitude at phase 0.5 when the mass accretion rate is relatively small ($\dot{m}=100$ and 1000). 
However, an inversion of the flux happens when $\dot{m}$ is large. 
This is because the optical $V$-band flux from an accretion disk increases as the mass accretion rate increases. 
This feature may be applicable to the observation data of SS433. 

In figure \ref{fig:beta2}, we change the wind velocities with various values. 
The size of the wind photosphere depends on the velocity of the wind 
 $\beta$.  
That is, the last scattering surface is inversely proportional to the velocity (see equation (\ref{eq:rph2})). The low $\beta$ outflow therefore makes large photosphere, and thus the geometrical thickness of the wind causes the deep secondary minimum during its eclipse. 
These features are the main results of the present study. 

\section{Discussion}

\subsection{Eclipsing Light Curves in SS433}

One difficulty of the optical light curves in SS433 is interpretating of the secondary minimum at phase 0.5, which is made by an eclipse of the companion star by the disk. 
As we explained in the previous section, it is difficult to fit the optical light curves observed in SS433 with only the disk model. 
The theory also support the massive outflow scenario introduced by observations.
The mass accretion rate in SS433 is appreciably supercritical, $\dot{M} \sim 10^{-4} M_\odot$/yr (van den Heuvel 1980; Shklovskii 1981; Perez \& Blundell 2009 ). This value is corresponds to $\sim 4.5 \times 10^7 \dot{m}$ for a $10 M_\odot$ black hole, and it seems to be an upper limit of the mass accretion rate. 

Since its discovery, binary parameters in SS433 have not yet been confirmed. 
In particular, we could not reach consensus as to whether the compact object in SS433 is a neutron star or a black hole. 
Gies et al. (2002) evaluated the mass of the companion star via absorption lines of A7Ib star to $19 \pm 7 M_\odot$. 
Recently Kubota et al. (2009) reported a new constraint on the mass of the compact object by the absorption lines taken from SUBARU and Gemini observations. They derive the mass of the compact object and companion star to be $M_{\rm X} = 4.1_{-0.7}^{+0.8} M_\odot$ and $M_{\rm C} = 12.2_{-2.1}^{+0.8} M_\odot$. 
%
%Recently a spectral data with high quality, however, the spectral type of the star which becomes a mass donor is proved and the mass with compact star is ending up with the constant value $q = M_{\rm X}/M_{\rm C} \sim 0.3-0.5 $ (Gies et al. 2002; Hillwig et al. 2004). 
%
We apply the mass ratio $q = M_{\rm X}/M_{\rm C} = 0.38$ derived by Kubota et al. (2009), and the black hole mass $M_{\rm BH}$ set to be $4.0 M_\odot$. 
This mass ratio is not in conflict with other observation results (Gies et al. 2002; Hillwig et al. 2004). 

\begin{figure}
  \begin{center}
    \FigureFile(60mm,60mm){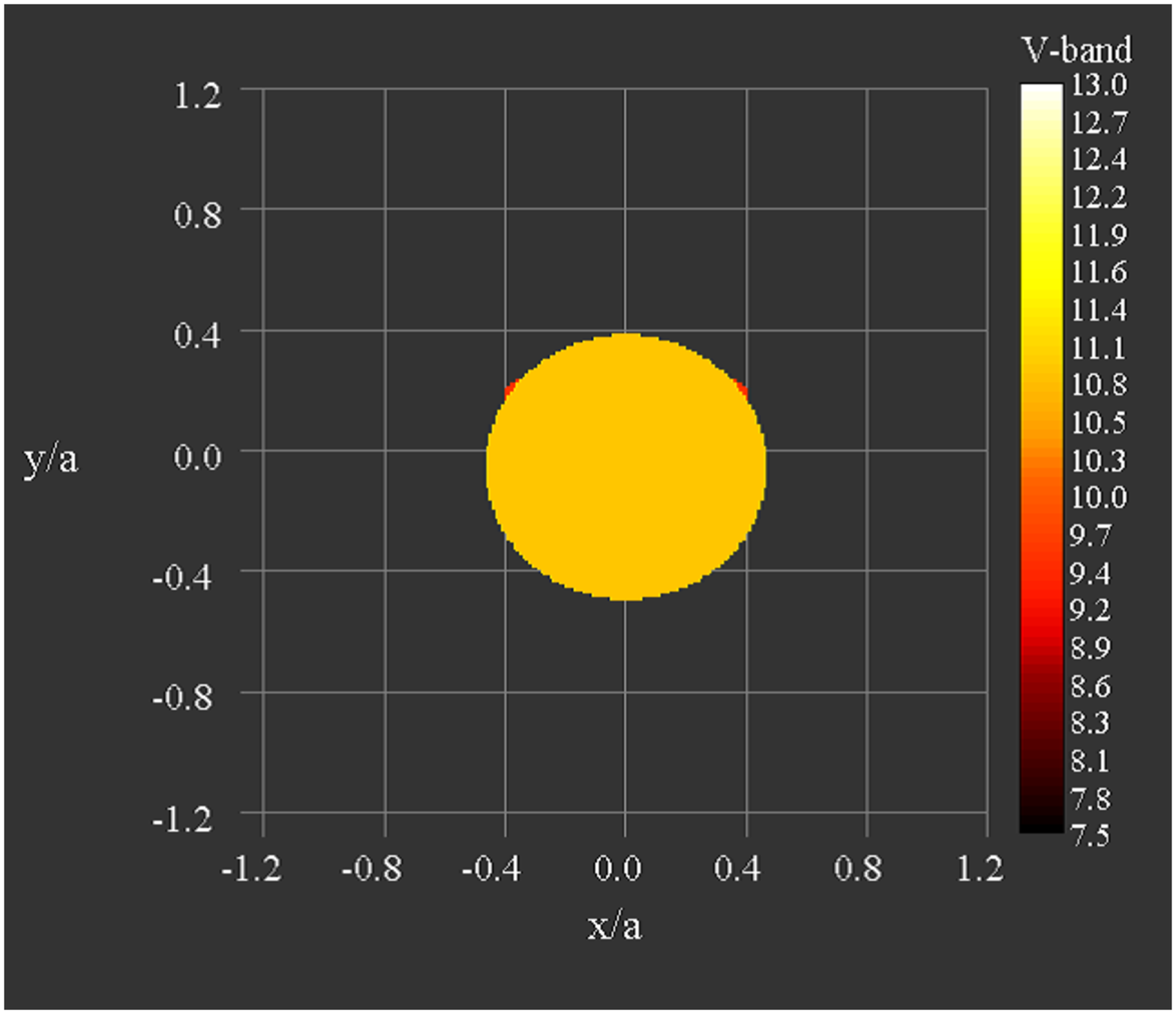}
    \FigureFile(60mm,60mm){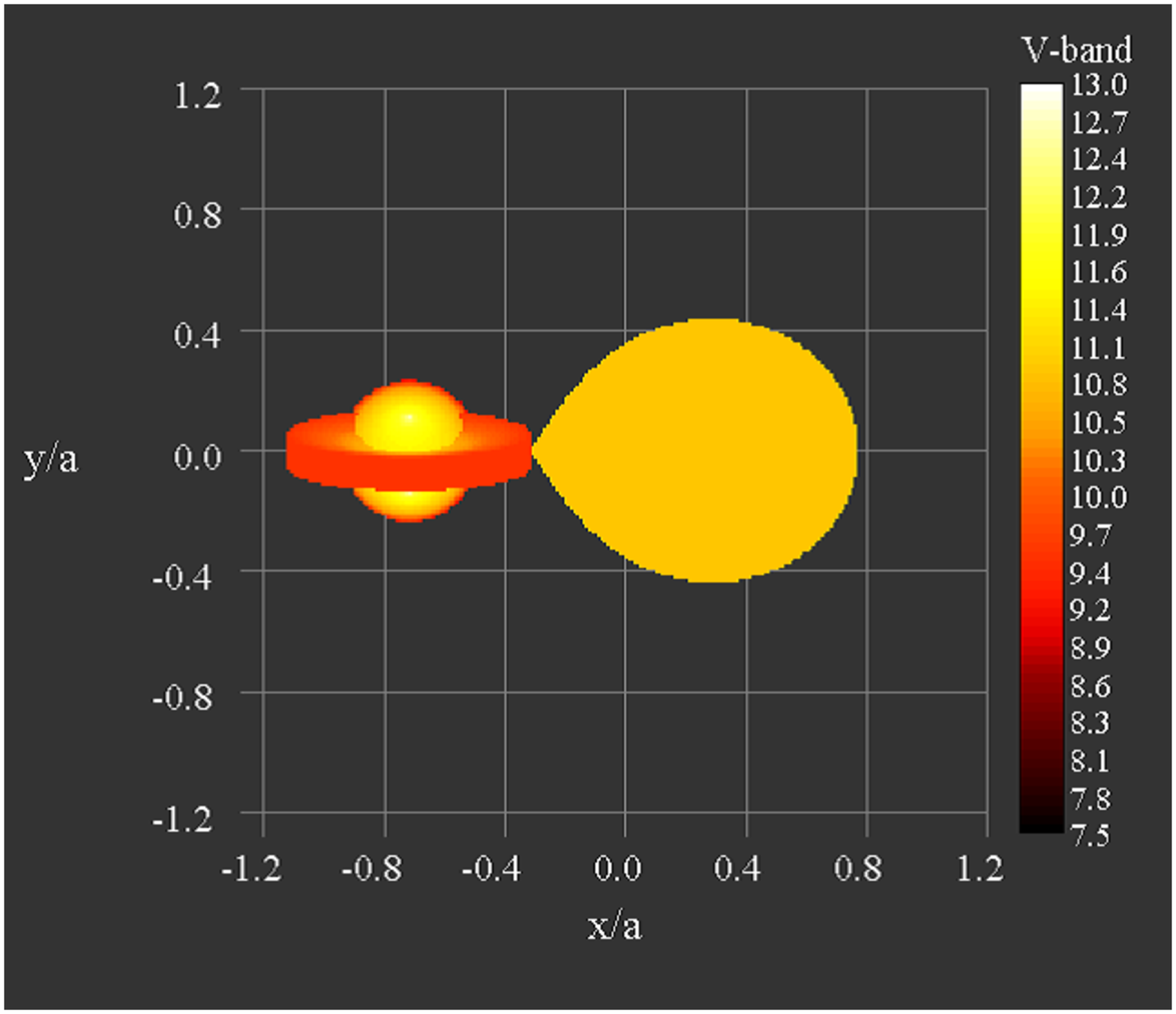}
    \FigureFile(60mm,60mm){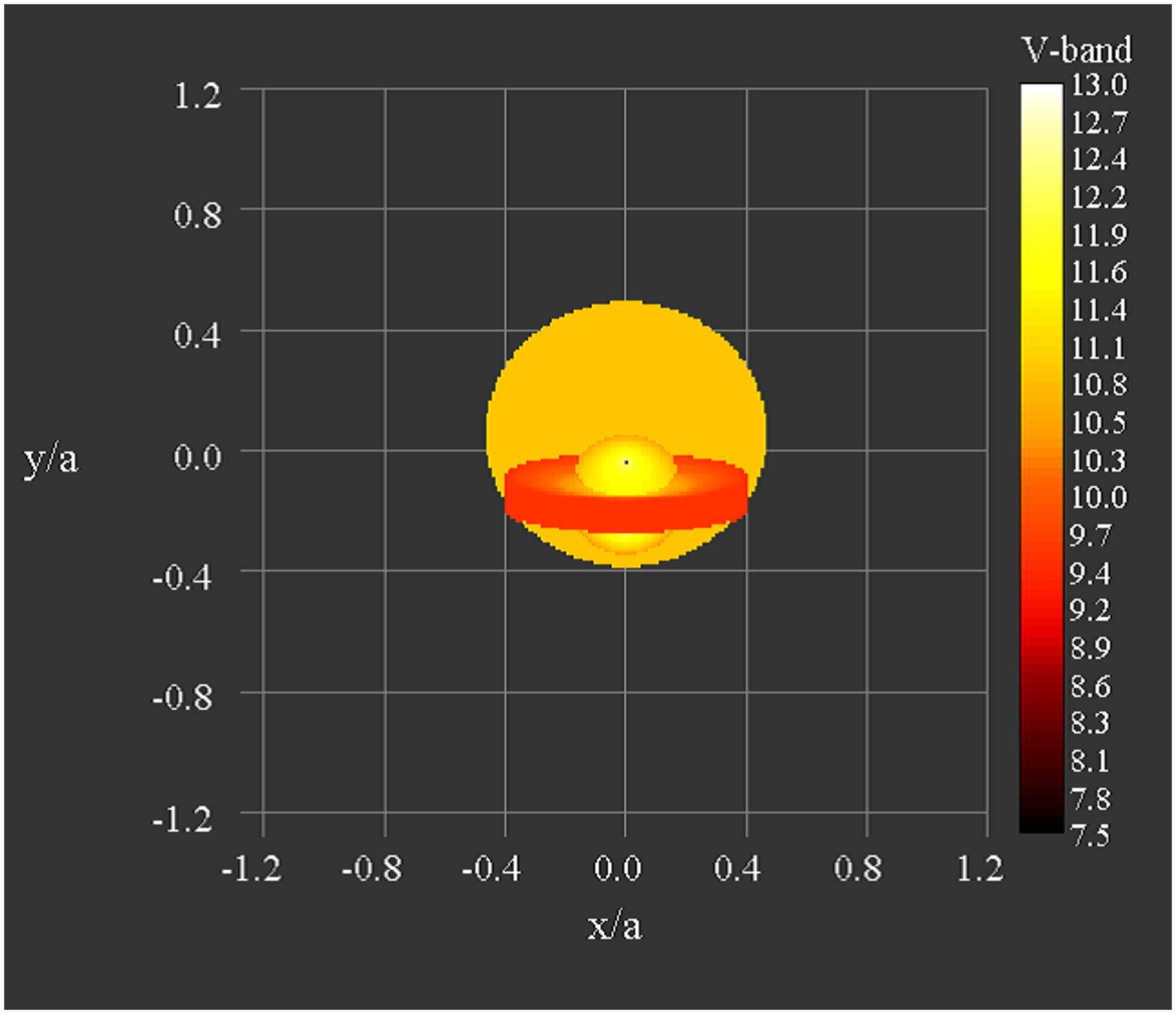}
    \FigureFile(60mm,60mm){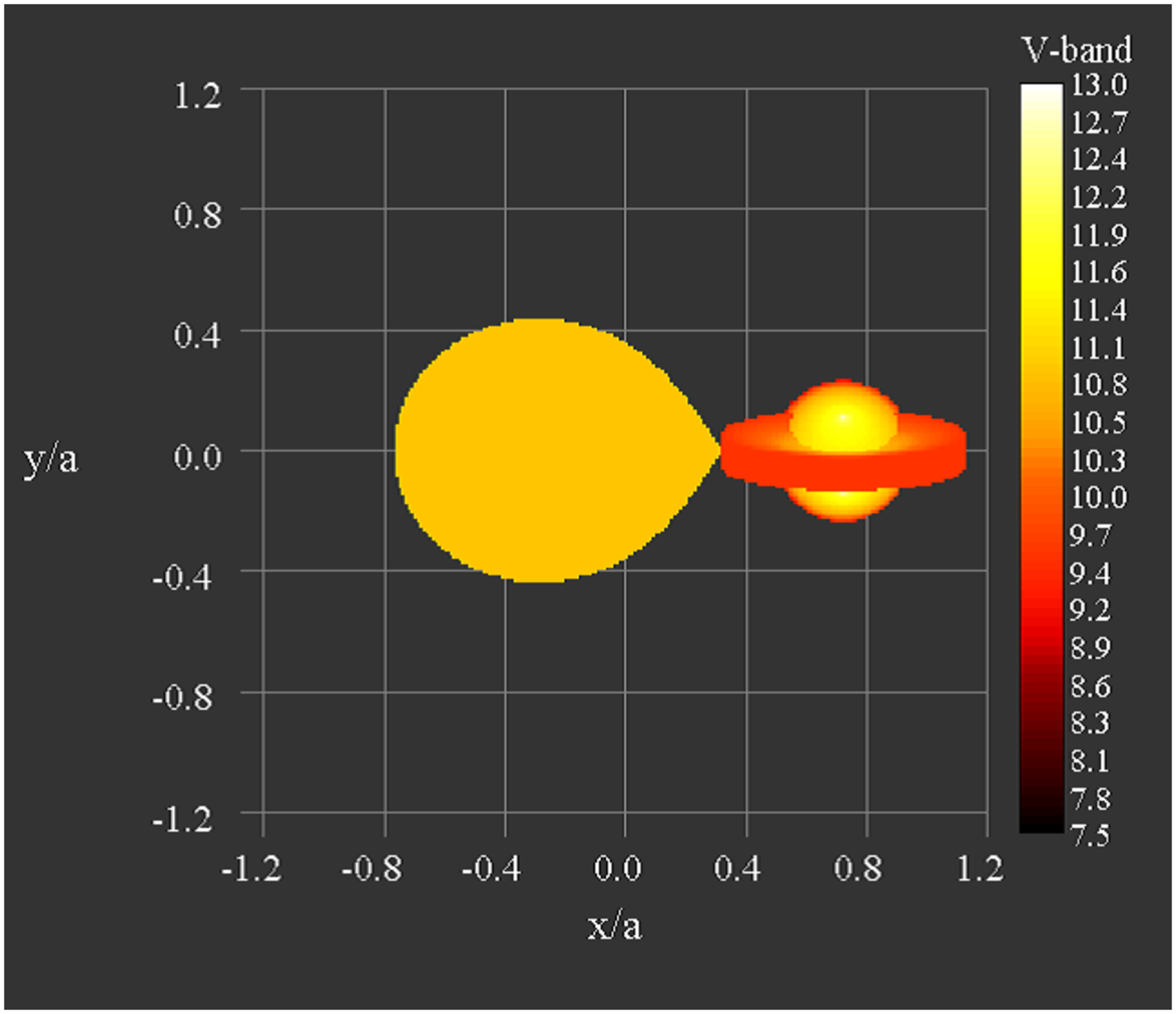}
    %%% \FigureFile(width,height){filename}
  \end{center}
  \caption{$V$-band flux image in SS433 at various phases (0, 0.25, 0.5, 0.75).
The mass accretion/outflow rate is $\dot{m}_{\rm acc}=\dot{m}_{\rm out}$=5000. 
The black hole mass is $M_{\rm BH}=4 M_{\odot}$, the mass of the companion star is $M_{\rm C}=12 M_{\odot}$, i.e., mass ratio $q=M_{\rm BH}/M_{\rm C}=0.38$. 
The effective temperature of the companion star is $T_{\rm C}=$15000 K, 
 and the inclination angle is $i=78^{\circ}$, whose values are referred from observational results. The velocity of the outflow is fixed at $\beta=0.01$. 
}
\label{fig:imgss433}
\end{figure}

Figure \ref{fig:imgss433} represents $V$-band images at various binary phases. 
We fix the mass ratio $q = 0.38$, inclination angle $i=78^\circ$, and the binary period $P=13.1$ day based on many former observations. 
The disk size (radius) is $1.35 \times 10^6 r_{\rm g} \sim 1.6 \times 10^7$ km, which is smaller than the case of $q=1.0$. 
The size of the photosphere is comparable to the disk size in the case of $\dot{m}$=5000. 

As for the wind velocity, a quasi-spherical non-relativistic wind from accretion disk has 3000 ${\rm km/s} \approx 0.01 c$ (Cherepashchuk 2002). 
Recently, Perez et al. (2009) found very fast accretion disk wind by near-IR spectroscopy, and its terminal velocity is about 1500 ${\rm km~s^{-1}}$ which is equivalent to 0.5 \% of the speed of light. 
In the fit, we fixed the wind velocity to $\beta=0.01$ for simplicity, allowing  changes the other two parameters, namely the mass accretion rate and temperature of the donor star. 

\begin{figure}
  \begin{center}
    \FigureFile(80mm,80mm){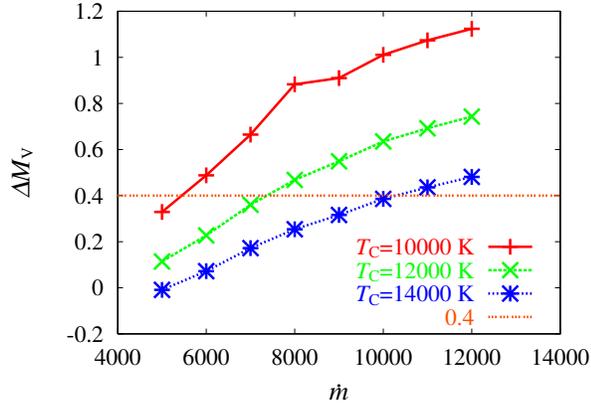}
    %%% \FigureFile(width,height){filename}
  \end{center}
  \caption{ Difference of $V$-magnitude in phase 0 and phase 0.5 as a function of mass accretion rate. Other parameters are fixed at $M_{\rm BH}=4 M_{\odot}, M_{\rm C}=12 M_{\odot}$, $\beta=0.01$, and $i=78^{\circ}$, respectively.  }
\label{fig:ss433vmag}
\end{figure}

\begin{figure}
  \begin{center}
    \FigureFile(80mm,80mm){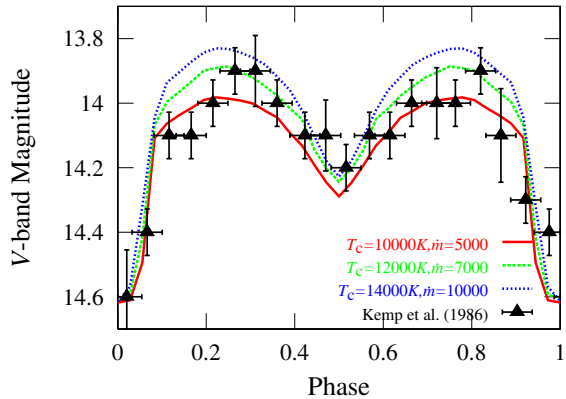}
  \end{center}
  \caption{$V$-band light curve of SS433 fitted by our model with several parameter sets. Kemp et al. (1986) showed the averaged optical light curve data of SS433. We set the numerical data to agree to the observed flux at phase 0. 
Other fixed the parameters are the same as figure \ref{fig:ss433vmag}. }
\label{fig:ss433fit3}
\end{figure}

Before comparing the observational data in SS433 and our model directly, we infer the possible parameter sets with figure \ref{fig:ss433vmag}. 
Figure \ref{fig:ss433vmag} represents the difference of $V$-magnitude ($\Delta M_{\rm V}$) at phase 0 and phase 0.5 as a function of accretion rate. 
The definition of the $\Delta M_{\rm V}$ is given by 
\begin{equation}
\Delta M_{\rm V} = M_{\rm V} ({\rm @phase~0.5}) - M_{\rm V} ({\rm @phase~0}). 
\end{equation}
In the case of SS433, $\Delta M_{\rm V}$ is about 0.4, which shows the magnitude of phase 0.5 is larger than that of phase 0. 
Actually, there are some parameter sets that can fit the observation, 
 we thus could not uniformly determine the best-fit parameter. 
We tried to fit our model to the $V$-band average light curve by Kemp et al. (1986). It is clear in figure \ref{fig:ss433fit3} that our model can fit whole shape of the light curve, but it does not make a distinction between three cases, i.e., (i) $\dot{m}$=5000, $T_{\rm C}$=10000K, (ii) $\dot{m}$=7000, $T_{\rm C}$=12000K, and (iii) $\dot{m}$=10000, $T_{\rm C}$=14000K, respectively. 

The mass-donor (companion) star is supposed to be A-type evolved star from the absorption line analysis (Gies et al. 2002; Hillwig et al. 2004). 
%It can be expected that the temperature of the companion star is therefore in the range of $T_{\rm C} = 17000-20000$ K. 
On the other hand, some recent observations indicate that the companion star has lower temperature which is less than 15000 K (Kudritzki et al. 2003; Cherepashchuk et al. 2005). 
It is an important task for observers to confirm the temperature of the companion star in SS433.

SS433 has a precession period of the jet,
 and the light curve changes at each precession phases. 
It is necessary to compare the light curve of each precession phase
 with the model testing. 

In addition, X-ray emission in SS433 comes from not only the photosphere of the wind but also the non-thermal emission of jet component (Rose 1995; Krivosheyev et al. 2009; see also Abolmasov et al. 2009). 
X-ray emission from the outflow depends on the acceleration mechanism,
 the temperature distribution, and the initial velocity of the outflow
launched from the disk surface. 
If we consider the X-ray emission more seriously, non-thermal X-ray emitting component (maybe jet component) should be included in our model (e.g., Reynoso et al. 2008; Cherepashchuk et al. 2009). 
These issues will be our future tasks. 

\subsection{Comments on Eclipsing ULXs}

Considering the probability of the eclipse events in the binary system, 
 it is natural that a few eclipsing binaries exist (Pooley \& Rappaport 2005). 
Recently X-ray eclipsing light curves has been detected in several ULXs. 
As the data of the ULXs increases, 
 the number of eclipsing ULXs also increases. 

M33 X-7 is one example of an eclipsing black hole X-ray binary discovered in recent years, and X-ray eclipse has been clearly detected (Pietsch et al. 2004). 
This object suspects a high mass X-ray binary, but its luminosity is very large unlike the famous Cygnus X-1. 
There are several scenarios to explain the high luminosity. 
Moreno M\'endez et al. (2008) pointed out the contradiction of the black hole spin scenario and proposed a hypercritical accretion scenario in M33 X-7 based on the binary evolution theory. 
An eclipsing luminous X-ray binary in the dwarf starburst galaxy NGC4214 has been reported by Ghosh et al. (2006). 
They clearly detected the X-ray eclipsing feature from this object. 

If both optical and X-ray eclipse are observed in an object, 
 we may presume the spatial structure of the accretion disk with the difference of the emitting region. 
We are looking forward to waiting for further detection of the eclipse in black hole binaries not only in our galaxy, but also in galaxies. 

\section{Conclusion}

We have calculated the light curves of eclipsing
 binaries by handling more realistic accretion disk models with an optically thick outflow.
We also applied the present model to the supercritical accreting
object SS~433. 
Our calculation is somewhat simple, but the geometrical thickness of the accretion disk has not been considered seriously so far. 
In addition, we clearly show the change of the shape of the light curve by the wind using a numerical calculation. 

As for the model of wind, because it includes unknown physics (accretion, geometrical thickness, mass loss rate, etc.), so we need to evaluate the temperature
 at the photosphere more seriously. 
This will be an important issue for observed PG quasars
 (Young et al. 2007). 
Modeling of the detailed physics of outflow is our future issue.

The measurement of the wind velocity and
 the mass loss rate via observations of the absorption lines will be crucial evidences to confirm our scenario. 
The fitting with the multi-wavelength observational data will be the next step to confirming our model. It is also one of our future tasks. 

%\acknowledgments
\vspace{10mm}
We would like to thank S. Mineshige for stimulating discussions. 
We also would like to thank K. Kubota for her helpful comments from the observational viewpoint. 
The author also would like to thank T.Suzumori for checking the manuscript. 
This calculation was supported by Kongo system of the Osaka Kyoiku University Information Processing Center. 
This work was supported by the Grant-in-Aid for the Global COE Program ``The Next Generation of Physics, Spun from Universality and Emergence'' from the Ministry of Education, Culture, Sports, Science and Technology (MEXT) of Japan. 

%This work was supported in part by the Grants-in Aid of the
%Ministry of Education, Science, Sports, and Culture of Japan
%(16004706, KW).

%%%%%%%%%%%%%%%%%%%%%%%%%%%%%%%%%%%%%%%

%\appendix
%%%
% See the manual for the detail.
%%%

\end{document}